**Title**: Impact of degree truncation on the spread of a contagious process on networks

**Running head**: Degree truncation and contagious processes on networks

**Authors**: Guy Harling,[1] Jukka-Pekka Onnela.[2]

1. Department of Global Health and Population, Harvard T.H. Chan School of Public Health
2. Department of Biostatistics, Harvard T.H. Chan School of Public Health

**Email addresses**: gharling@hsph.harvard.edu; onnela@hsph.harvard.edu

**Corresponding author**: Jukka-Pekka Onnela, Department of Biostatistics, Harvard T.H. Chan School of Public Health, 677 Huntington Avenue, Boston, MA 02115.

**Acknowledgements**: We thank members of the Onnela lab for feedback on earlier versions of this paper. This research was supported by P30 AG034420.

**Abstract**

Understanding how person-to-person contagious processes spread through a population requires accurate information on connections between population members. However, such connectivity data, when collected via interview, is often incomplete due to partial recall, respondent fatigue or study design, e.g., fixed choice designs (FCD) truncate out-degree by limiting the number of contacts each respondent can report. Past research has shown how FCD truncation affects network properties, but its implications for predicted speed and size of spreading processes remain largely unexplored. To study the impact of degree truncation on spreading processes, we generated collections of synthetic networks containing specific properties (degree distribution, degree-assortativity, clustering), and also used empirical social network data from 75 villages in Karnataka, India. We simulated FCD using various truncation thresholds and ran a susceptible-infectious-recovered (SIR) process on each network. We found that spreading processes propagated on truncated networks resulted in slower and smaller epidemics, with a sudden decrease in prediction accuracy at a level of truncation that varied by network type. Our results have implications beyond FCD to truncation due to any limited sampling from a larger network. We conclude that knowledge of network structure is important for understanding the accuracy of predictions of process spread on degree truncated networks.

**Introduction**

Our understanding of how disease, knowledge and many other phenomena spread through a population can often be improved by investigating the population's social or other contact structure, which can be naturally conceptualized as a network (Newman, 2002; Pastor-Satorras et al., 2015). In the case of human populations, this contact structure is often gathered through the use of questionnaires or surveys that typically ask respondents to name some of their contacts (Burt, 1984; Holland & Leinhardt, 1973). Generating population-level network structures from such data requires one of two possible approaches (Marsden, 2005). One approach is to delineate a population of interest, interview every person in the population, and collect unique identifiers for each respondent's contacts; this allows the mapping of the true *sociocentric* network within that population. The alternative is to sample the population of interest and collect information about each respondent and his or her contacts; this results in a collection of *egocentric* networks from that population. Either approach enables the extraction of network features that can be used to fit a graph model, such as one of the models in the family of exponential random graphs (ERGMs) (Lusher et al., 2012), which allows the subsequent generation of network graphs consistent with the fitted features of the observed networks. The features that may be extracted from egocentric networks are however quite limited, making sociocentric networks the preferred design, resources allowing.

Both egocentric and sociocentric approaches can place a considerable burden on the respondent to recall numerous contacts and describe each in detail (McCarty et al., 2007). As a result, most sample survey questionnaires, in both egocentric and sociocentric designs, limit the contacts sought from a respondent, for example by the content, intimacy level, geographic location or time frame of the relationship elucidated (Campbell & Lee, 1991). A common method is to limit the number of contacts a respondent describes. This may be done directly, e.g. by asking "who are your five closest friends with whom you regularly socialize?" It may also be done indirectly, e.g. by asking "who are the friends with whom you socialize" but then only asking follow-up questions about the first five named (Burt, 1984; Kogovsek et al., 2010). A less-common variant of the second approach is for the interviewer to ask follow-up questions on a random subset of named contacts.

More recently, there has been increasing interest in leveraging large-scale data on digitally mediated social interactions ranging from emails to online social networking services to mobile phone communication. Call detail records, resulting from mobile phone calls and text and multimedia messages, have become especially popular for capturing one-to-one social interactions in large populations (Blondel et al., 2015; Onnela et al., 2007a; Onnela et al., 2007b). Social networks are constructed from these data typically by aggregating longitudinal interactions over a time window of fixed length, where the features of the resulting networks are fairly sensitive to the width of the aggregation window (Krings et al., 2012). Although some *ad hoc* approaches have been proposed, so far there are no statistically principled methods available for setting the window width. This leads to effective network degree truncation that is not a consequence of study design per se but rather is induced by the network construction process. While our focus here is degree truncation resulting from study design, we point out that many of our findings are applicable in other settings as well.

All of the above approaches potentially lead to truncation of the number of observed contacts. There is longstanding concern within the sociological literature that such truncation might affect estimates of network properties, including various forms of centrality (Holland & Leinhardt, 1973). However, there are countervailing resource and data quality benefits to avoiding respondent and interviewer fatigue (McCarty et al., 2007). While investigating the effect of degree truncation on structural properties of networks is an important problem, substantive interest often lies in making inferences about how a dynamical process on the network, such as epidemic spread, might be affected by truncation. Surprisingly, while both the impact of degree truncation on structural properties of networks and the impact of structural properties on the spread of a dynamic process through a networked population have been investigated, the joint implications of the two processes have not yet been elucidated. To integrate key ideas from the two corpora, we review first the literature on the impact of truncating reported contacts on structural network properties, and second the literature on the impact of structural network properties on spread dynamics, to arrive at hypotheses regarding how truncation might change expected spreading process outcomes. While our work was motivated by epidemic disease processes, our analysis should be applicable to any process that can be modeled using compartmental models of epidemic spread. We test the predictions of our hypotheses with simulation models using both synthetic, structured networks, and empirically observed networks.

Spreading processes on networks can be modeled by generating ensembles of networks weighted by their similarity to observed egocentric data on certain key structural features, e.g., using Exponential Random Graph Models (Lusher et al., 2012), or in a Bayesian framework the networks can be sampled from an estimated posterior distribution of networks consistent with the observed data (Goyal et al., 2014). Once the graphs have been generated, a series of dynamic processes are run across this collection of graphs (Jenness et al., 2015). However, using this modeling approach to explore the impact of truncation would conflate two processes: the truncation process and the network generation process. In order to focus on the former, we generate multiple realizations of synthetic full-network datasets with specific network properties, and additionally utilize a collection of empirically observed sociocentric networks that can be interpreted as multiple network realizations from a larger meta-population. As a result, we are able to isolate the effect of degree truncation and explore its impact on spreading processes on networks with very different structural properties.

*The impact of contact truncation on structural network properties*

Limiting the number of connections ("alters") reported by a respondent ("ego") in both egocentric and sociocentric designs is known as a *fixed choice design* (FCD) (Holland & Leinhardt, 1973). This limitation right-censors (imposes an upper bound on) an ego's *out-degree* (the number of alters nominated by an ego). In sociocentric studies out-degree truncation may in turn reduce the *in-degree* of others, because some existing ties may end up unreported due to the constraints on out-degree. (In directed networks the mean in-degree and mean out-degree must match, since each edge adds one to the out-degree of the "source" node and one to the in-degree of the "sink" node.) Thus in the corresponding undirected network, obtained from the directed network simply by ignoring the directions of the edges and considering a mutual or two-way nomination as a single edge, the truncated (observed) total degree (either in or out) will be:

$$k_i^t = \begin{cases} k_i \text{ if } k_i \leq k_{fc} \\ k_{fc} \text{ otherwise} \end{cases}$$

where $k_{fc}$ is the FCD truncation value and $k_i^t$ each individual's degree in the truncated network graph. For node $i$ with $k_i > k_{fc}$, edge $e_{ij}$ between ego $i$ and alter $j$ will be maintained for certain only if $k_j < k_{fc}$, otherwise ego $i$ must nominate alter $j$ in order for the edge to be observed. FCD can be conducted in two ways, as outlined above. The more-common approach of focusing on the first $k_{fc}$ or fewer names reported ("weighted truncation") is likely to lead to bias towards stronger contacts, since stronger ties are likely to be more salient to a respondent. This approach should thus maximize the proportion of a respondent's social interactions that is captured. The less-common approach of drawing a random subset of all named contacts ("unweighted truncation") will provide a broader picture of the types of contacts a respondent has – notably increasing the chance of observing weak ties – at the cost of observing a smaller proportion of the respondent's total social interaction. FCD is known to impact several canonical network characteristics, but its effects depend on the structure of the complete network graph (Kossinets, 2006); we consider next some key properties.

*Degree distribution and assortativity.* While the impact of FCD on the *network degree distribution $p_k$* is almost always to reduce its mean and variance, its precise effect depends on both the first and second moment of the degree distribution and on the ratio of $k_{fc}$ to the mean degree $\mu_k$. Lower $k_{fc}/\mu_k$ ratios generally increase (strictly, never decrease) the number of edges dropped. Networks with higher-variance degree distributions will also typically lose more edges for a given $k_{fc}/\mu_k$ ratio, insofar as such networks have a larger proportion of nodes with degree greater than $k_{fc}$, and thus for any given edge $e_{ij}$ both $k_i$ and $k_j$ are at higher risk of being truncated, leading to the edge $e_{ij}$ being dropped.

Loss of edges in high-variance networks may, however, be offset by degree-assortativity (Kossinets, 2006), often quantified by the Pearson correlation coefficient of degrees of connected nodes: $r = \frac{\Sigma_{xy}(e_{xy}-a_x b_y)}{\sigma_a \sigma_b}$, where $e_{xy}$ is the fraction of all edges that join nodes of degree $x$ and $y$, $a_x$ and $b_y$ are the fraction of edges that start and end, respectively, at nodes of degree $x$ or $y$, respectively, and $\sigma_a$ and $\sigma_b$ are the standard deviations of distributions of $a_x$ and $b_y$ (Newman, 2003b). If the network is degree-disassortative, such as the scale-free Barabási-Albert network where $p_k \sim k^{-\gamma}$ and $2 < \gamma < 3$ (Barabási & Albert, 1999), then edges that might be censored by the adjacent high-degree node are less likely to also be censored by the adjacent low-degree node, and thus dropped entirely in the truncated network (Vázquez & Moreno, 2003). Degree-assortative, high-variance networks are thus likely to see the greatest change in $p_k$; human contact networks are typically somewhat degree-assortative, and while communication contacts have fat-tailed degree distributions with high variance, physical contact networks are more degree-homogeneous (Onnela et al., 2007a; Onnela et al., 2007b; Salathé et al., 2010). The level of degree-assortativity in a network is not itself systematically affected by FCD, so long as edges are dropped without regard to the strength of each connection (Kossinets, 2006; Lee et al., 2006). However, if individuals are more likely to report stronger connections, and ties between individuals of similar degree are more likely to be strong – which is suggested by the combination of findings that homophilous ties are more likely to be transitive (Louch, 2000; Marsden, 1987) and those with greater transitivity (Onnela et al., 2007b) tend to be stronger – then FCD might be expected to artificially inflate $r$.

*Clustering*. Local clustering can be measured in at least two different ways: (i) *Triadic clustering*: the mean of local clustering coefficient $C_i$, where $C_i$ is the ratio of the number of ties present between all neighbors of node $i$ and $k_i(k_i - 1)/2$, the number of pairs of neighbors of $i$ (Watts & Strogatz, 1998); (ii) *Focal clustering*: the level of global triadic closure, that is the ratio of triangles – where $(u, v)$, $(u, w)$ and $(v, w)$ are all present – to paths of length two, i.e., if $(u, v)$ and $(v, w)$ exist, they form a path of length two (Newman, 2010). Clustering may also occur at higher levels of aggregation in the network, for example in the presence of network communities where, loosely speaking, the density of edges within a set of nodes belonging to a community is higher than the average density of edges across the whole graph (Fortunato, 2010; Porter et al., 2009). One way to quantify this community-level clustering is by modularity, $Q = \sum_r(e_{rr} - a_r^2)$, where $e_{rr}$ is the proportion of edges in the network that connect nodes in community $r$ to other nodes in community $r$ and $a_r$ is the proportion of ends of edges that are attached to nodes in community $r$ (Newman, 2006). The value of modularity can be normalized using the degree distribution of the network as $Q_n = Q/[1 - \sum_r(k_i k_j/2m)\delta(c_i, c_j)/2m]$, $m$ is the number of edges in the network and $\delta(c_i, c_j)$ is equal to one if $c_i = c_j$ and zero otherwise. This normalization makes modularity values more readily comparable across networks (Newman, 2010).

When truncation is unweighted, we expect FCD to reduce clustering at the triadic and community levels as it effectively results in random edge removal. When truncation is weighted, however, FCD might lead to an increase in clustering: if within-cluster edges are stronger than others, they are more likely to be preserved.

*Path lengths*. In removing ties, unweighted FCD will reduce the fractional size of the *largest connected component* (LCC), $S_{LCC}$, and will often increase the *average path length between nodes of the LCC*, $\ell_{LCC}$. If FCD is weighted, however, the rate at which $\ell_{LCC}$ falls may be reduced, at the expense of a faster decline in $S_{LCC}$, as peripherally (weakly) connected nodes are preferentially dropped from the LCC. In a network with a dense core, the $S_{LCC}$ is likely to be better preserved in a degree-disassortative than in a degree-assortative network under FCD – due to the lower probability of ties within the core being dropped from both ends (Kossinets, 2006). This effect will be more pronounced if the ties within this core are also stronger than other ties, and thus more likely to be preserved. For some specific threshold value of FCD the LCC will fragment, and two previously connected nodes may become disconnected.

While the above discussion, as is common in network characterizations, considers shortest paths, random spreading processes, like random walks, rarely take the shortest path from between two given nodes $i$ and $j$. Because of this, the length of the shortest path between nodes $i$ and $j$ in a fully observed network typically underestimates the length of the path taken by a stochastic spreading process. Partial observation of the network, such as that induced by degree truncation, inflates the lengths of shortest paths, but does of course not alter the length of the actual unobserved paths taken by the spreading process. For this reason, perhaps somewhat paradoxically, shortest paths inferred from partially observed networks can provide more accurate predictions of the path lengths taken by spreading processes than those based on fully observed networks (Onnela & Christakis, 2012).

*The impact of structural network properties on spreading processes*

There is a burgeoning literature on the effect of various network properties on spreading process outcomes (Barrat et al., 2008; Newman, 2002; Pastor-Satorras et al., 2015). Under assumptions of population homogeneity, relatively simple solutions can be found for key properties; however these results rarely hold once we allow for any non-trivial network structure (Keeling & Eames, 2005). We first outline canonical results under homogeneity, and then consider how structural network properties impact spreading processes. We focus on two aspects of an outbreak: the early stage and the final state. To simplify our analysis, we follow the tradition in this literature and focus on models that assume degree infectivity, where an infectious individual can infect all their neighbors in just one time step, rather than unit infectivity, where they can only infect one of their neighbors per time step (Staples et al., 2015).

In the early stages, we are typically interested in two related quantities: the *basic reproduction number* $R_0$, the number of new incident cases (newly infected individuals) arising from each currently infected individual in a fully-susceptible population; and $r_0$, the initial exponential (or faster) *growth rate* of an epidemic. $R_0$ is defined in all settings as a function of $\beta = \tau c$, where $\tau$ is the probability of infection per period and $c$ the number of contacts per period, and the rate at which individuals recover (or die), $v = 1/D$, where D is the mean duration of infectiousness. In a homogeneous mass-action model for an infection where recovery leads to immunity, i.e. a Susceptible-Infected-Recovered (SIR) model, $R_0 = \beta/v$, where $R_0 \geq 1$ ensures a large outbreak with non-zero probability (Hethcote, 2000). $r_0$ is conceptually equal to $\beta$ in the first period, but thereafter is not well-defined analytically – even in homogenous models – and is typically measured empirically as the second moment of the epidemic curve in its initial growth phase (Vynnycky & White, 2010). At the end of an outbreak, we can evaluate its overall impact via the *attack rate* $A = \Sigma_t I_t/N$, the proportion of the population ever infected.

*Degree distribution and assortativity.* In a network setting, $R_0$ can be viewed as the average number of edges through which an individual infects their neighbors across the whole period of their infectiousness if all their neighbors are susceptible. The probability of infection, $\lambda$, for each node can be conceptualized (for degree infectivity) in terms of their degree and their neighbors' infection statuses. In a degree-homogenous network, an epidemic will probabilistically take off if $\lambda(k) \equiv \lambda\langle k \rangle \geq 1$; when networks are degree-heterogeneous, the likelihood of epidemic take-off becomes a function of the first and second moments of the degree distribution (Newman, 2002). In an infinitely large scale-free network where $2 < \gamma \leq 3$, $\langle k^2 \rangle \to \infty$, and thus an epidemic occurs in all cases, regardless of the infection and recovery parameters (Barthelemy et al., 2005; Boguñá et al., 2003; May & Lloyd, 2001; Pastor-Satorras & Vespignani, 2001). More generally, higher degree heterogeneity increases $R_0$.

Similarly, higher degree-assortativity increases the chances of epidemic take-off. On finite networks the probabilistic threshold for epidemic take-off has a lower-bound of $\langle \bar{k}_{nn} \rangle$, the average degree of nearest neighbors, which is also the driver of both degenerate results: $\langle \bar{k}_{nn} \rangle = \langle k \rangle$ in a homogeneous network and $\langle \bar{k}_{nn} \rangle \to \infty$ in an infinitely large scale-free network (Boguñá et al., 2003). This is intuitive, since the number of one's neighbors bounds the number of infections one can generate.

The speed of epidemic growth is closely related to $R_0$ and $\lambda$. On an infinite size, scale-free network, $r_0$ is extensive, since once infection reaches high-degree nodes it spreads to a finite fraction of persons at the next time point (Vazquez, 2006a). More generally, an epidemic on any scale-free network will see growth at a power law rate such that early in the epidemic infection levels will be greater than is predicted by homogeneous models, in which growth rates are exponential (Vazquez, 2006b).

More generally still, degree-assortativity has long been known to lead to faster epidemic take-off, but a lower final epidemic size, conditional on the number of nodes and ties within a network (Gupta et al., 1989). This result arises from a dense core of high-degree nodes in which infection is rapidly passed, in combination with longer paths to peripheral, low-degree nodes where chains of infection are more likely to die out.

*Clustering*. The most straightforward effect of triadic clustering, for a given degree distribution, is to reduce the average number of infections each infected person causes. This reduction is due to newly-infected individuals having fewer susceptible neighbors: the contact who infected you is likely also have had the opportunity to infect your other contacts (Keeling, 2005; Miller, 2009; Molina & Stone, 2012). This does not strictly imply a lower $R_0$, since $R_0$ refers to a completely susceptible population, however this phenomenon increases the epidemic threshold in the same manner that a fall in $R_0$ would (Molina & Stone, 2012). Similarly, the epidemic growth rate $r_0$ is somewhat slowed by this reduction in the proportion of susceptible alters (Eames, 2008).

In many networks, e.g. Erdős–Rényi graphs (Erdős & Rényi, 1959), for a given network density, increased clustering also leads to a smaller $S_{LCC}$, which necessarily reduces the maximum possible epidemic size (Newman, 2003a). However, within the LCC clustering increases the density of the network (Serrano & Boguñá, 2006), providing more local pathways from an infected to a susceptible individual. This reduces the protective effect of any alters who have recovered without infecting an ego, and thus some simulations have found clustering increases the attack rate *A* (Keeling, 2005; Newman, 2003a).

Overall, cliques alone appear to have marginal effects on epidemic dynamics, however the processes which drive clique formation – such as homophily by nodal attributes or geographic proximity – lead to networks displaying clustering that also contain other topological features – such as degree-assortativity or heterogeneity – which do significantly affect epidemics, leading to processes on clustered networks looking very different from those on non-clustered ones (Badham & Stocker, 2010; Molina & Stone, 2012; Volz et al., 2011). Broader community structure in networks acts in much the same fashion as cliques, reducing $r_0$ due to limited capacity to pass infection from one community to the next (Salathé & Jones, 2010); although epidemics are unhindered, or even sped up, by inter-community ties when overlapping, rather than distinctly separated, communities are built into networks (Reid & Hurley, 2011).

*Conjectured impact of degree truncation on spreading processes*

Based on the above results, we formulate some initial hypotheses about the likely impact of out-degree truncation on the behavior of spreading processes on the resulting network. First and foremost, truncation will reduce the number of edges in the network, since some edges are now not observed. This leads to a reduction in mean degree and is likely to increase average path lengths and reduce the size of the $s_{LCC}$; as a result, both $r_0$ and $A$ will be reduced. The reduction in $r_0$ may however be offset by reduced variance in degree – since out-degree variance is strictly reduced by truncation and in-degree variance is likely to drop too. Second, degree truncation by tie strength may lead to an inflation of degree-assortativity, if assortative ties are stronger on average and thus more likely to be preserved. This should lead to smaller, faster epidemics – especially if assortativity is created by preferentially dropping core-periphery links. Finally, degree truncation by tie strength will have an unpredictable effect on clustering – depending on the relationship between tie strength and community structure. Notably, if the two are strongly positively correlated, truncation may increase community structure as weak ties are preferentially dropped. If clustering is increased, both $r_0$ and A are likely to fall.

**Methods**

To test the above hypotheses about the impact of degree truncation on spreading process outcomes, we: (1) simulated a truncation process on a range of networks; (2) simulated a spreading process on the original (fully observed or 'full' network) and truncated networks a large number of times; and (3) compared epidemic outcome values for the full and truncated networks (Figure 1). In the following, we describe in detail the following: (A) the network generation process; (B) the truncation process; and (C) the spreading process.

*A. Network structures*

We considered four types of synthetic networks that we call Degree-Assortative, Triadic Clustering, Focal Clustering, and Power-Law networks, and in addition we considered networks based on empirical data (details below). For unweighted synthetic networks, we used edge overlap as proxy for tie strength, defined as the fraction of shared network neighbors of a connected dyad: $O_{ij} = n_{ij}/[(k_i - 1) + (k_j - 1) - n_{ij}]$, where $n_{ij}$ is the number of neighbors $i$ and $j$ have in common, and $k_i$ and $k_j$ are their degrees (Onnela et al., 2007b). Overlap was shown to be strongly correlated with tie strength, as conjectured by the weak ties hypothesis several decades earlier (Granovetter, 1973). The empirical social networks were collected in 75 villages in Karnataka, India, which were surveyed as part of a microfinance intervention study in 2006 (Banerjee et al., 2013a, 2013b). We defined an edge between two individuals to exist if either person reported any of the twelve types of social interaction asked about in the study.

We began synthetic network construction by generating a collection of degree sequences, where a degree sequence is a list of node degrees of a network. To generate 100 Degree-Assortative, Triadic Clustering, and Focal Clustering networks, each consisting of $N = 1000$ nodes, we drew 100 degree sequences of length $N$ from a Poisson distribution $P(\phi)$ where $\phi = 8$, as an approximation to a binomial distribution with large N. We used the configuration model to generate an initial graph realization for each degree sequence (Molloy & Reed, 1995), and then rewired the networks, edge by edge, in order to obtain a collection of calibrated networks such that each network closely matches a target value of a chosen characteristic, specifically:

1. *Degree-Assortative*. This was achieved by: (i) selecting two disjoint edges $(u, v)$ and $(x, y)$ uniformly at random; (ii) computing whether removing the two edges and replacing them with edges $(u, y)$ and $(x, v)$ would increase network assortativity; and if so (iii) making this change.

2. *Triadic Clustering*. This was achieved by: (i) choosing an ego $i$ and two of its alters, $j$ and $k$, who were not connected to one-another; (ii) adding the edge $(j, k)$ to the network, thus forming a triangle; and (iii) removing an edge selected uniformly at random conditional on that edge not being part of a triangle, thus ensuring increased triadic clustering.

3. *Focal Clustering*. This was achieved by: (i) selecting three nodes $i$, $j$ and $k$ uniformly at random; (ii) adding edges $(i, j), (i, k)$ and $(j, k)$ if they did not already exist; (iii) choosing uniformly at random in the network the same number of edges that were just added (excluding edges $(i, j), (i, k)$ and $(j, k)$ in the selection); (iv) computing whether removing this second set of edges would result in a net increase in focal clustering – if so, removing them; if not, repeating steps (iii) and (iv).

We generated three versions of each type of synthetic network by calibrating assortativity, triadic clustering, and focal clustering to the minimum, median and maximum values of these quantities observed in the 75 Karnataka villages (Table 1).

To generate Power-Law networks, the fourth type of synthetic network, we drew degree sequences from a power-law distribution $P(k) \sim k^{-\gamma}$, using the values 3, 2.5 and 2 for the degree exponent $\gamma$. We discarded any ungraphable sequences, i.e. those where any value greater than $N - 1 = 999$ was drawn. We again used the configuration model to generate an initial graph realization for each degree sequence. Note that lower values of $\gamma$ are associated with degree distributions that have increasingly fat tails.

For each of the four types of synthetic networks, for each level of calibration we generated 100 independent representative networks using the above methods, for a total of 1200 networks.

## *B. Truncation*

We simulated degree truncation of the form typically seen in surveys, by placing a ceiling on the number of contacts, $k_{fc}$, that can be reported by a respondent, and then reconstructed the contact graph created from all sampled contacts. To do this, we first converted the network into a directed graph. We then selectively removed $(k_i - k_{fc})$ directed edges starting from each individual $i$, beginning with the edge having the smallest edge overlap value (the "weakest" edge); we were thus conducting truncation by tie strength. We truncated at $k_{fc} = q\langle k \rangle$, taking values of $q = 0.5, 1, 2$, so that the maximum out-degree of individuals was half the mean degree in the full network, the same as its mean degree, or twice its mean degree. After truncating each individual's out-degree, we collapsed the directed graph into an undirected one based on all remaining ties. Examples of this truncation process on 20-node networks are shown in Figure 2. We measured a range of network properties for each full and truncated network, including mean degree, degree-assortativity, triadic and focal clustering, $s_{LCC}$ and normalized modularity $Q_n$ – this last based on a graph partition for each network using the Louvain method (Blondel et al., 2008).

*C. Spreading process*

We ran a Susceptible-Infected-Recovered (SIR) model on the networks defined by the per-period (per time step) probabilities $\beta = 0.03$ (the probability of an infectious individual infecting each susceptible contact) and $\nu = 0.05$ (the probability of an infectious individual recovering). Each spreading process began with five initial infections, chosen uniformly at random among the nodes of a network, and each SIR model was run 100 times on the full and degree truncated variants of each of the 100 networks. We measured two categories of outcomes across all of the 10,000 runs (100 runs per network for 100 networks) of each synthetic network type (7,500 for the Karnataka village data), including results from those runs for which at least 10% of individuals were ever infected: first, *time to infection* of the 10$^{th}$ percentile of the population (epidemic growth $r_0$: mean and 95% range); and second, the *proportion of nodes ever infected* (the attack rate $A$: mean and 95% range).

**Results**

Summary statistics for all networks at all levels of truncation are shown in Supplementary Table 1. In all networks, both synthetic and empirical, out-degree truncation consistently reduced mean degree as expected, most strongly in Power-Law and Focal Clustering networks. Truncation strongly reduced degree-assortativity in all cases except for Power-Law networks, which were already degree-disassortative, overwhelming any differences originally seen across levels of calibration; this effect was weaker for the Karnataka networks than for synthetic networks other than Power-Law. Modularity increased with truncation in all networks except for Degree-Assortative ones (which had very high initial modularity). With the exception of Power-Law and Karnataka networks, where modularity rose smoothly with increasing truncation, most of the increase only occurred once networks were truncated at half mean degree. Both triadic and focal clustering fell, and the $\ell_{LCC}$ rose, consistently with increasing truncation for all networks in which clustering was initially present.

When epidemics were simulated on the full networks, the attack rate $A$ was $\geq$10% in almost every simulation (over 97.5%), with the exception of Degree-Assortative networks where only around 90% of simulations reached $A \geq 10\%$ (Supplementary Table 2). Truncating networks at $2\langle k \rangle$ had almost no impact on the proportion of epidemics with $A \geq 10\%$ for any network, but further truncation led to a sharp fall-off. At $0.5\langle k \rangle$ truncation none of the clustered network epidemics reached $A \geq 10\%$, and only the Power-Law networks, the Degree-Assortative networks calibrated to the lowest level of assortativity and the Karnataka networks had more than 2% of their epidemic reach the $A \geq 10\%$ threshold.

Without truncation, 10% of all nodes were infected within 20 time steps on all networks except for the degree-assortative ones – which also showed the greatest range of $r_0$ (Table 2). Truncation at $2\langle k \rangle$ increased $r_0$ in all cases, but not by large amounts; however truncation at $\langle k \rangle$ both raised mean $r_0$ and its variance – notably in the cases of degree-assortative and triadic clustering networks (Figure 3A). For those networks in which any runs reached $A \geq 10\%$ at $0.5\langle k \rangle$ truncation, both the mean and variance of $r_0$ increased as networks became highly fractured. Network structure had a greater impact on $A$ than on $r_0$, with clear differences even on full networks (Figure 3B). Truncation at $2\langle k \rangle$ had almost no impact on $A$ except in the cases of Power-Law, and to a lesser extent Degree-Assortative, networks. However truncation at $\langle k \rangle$ leads to a mean $A$ roughly halving for all cases except the Karnataka networks, where $A$ only falls by about a quarter. Once truncation reached $0.5\langle k \rangle$, no network type averaged $A > 16\%$.

**Discussion**

Simulating a generic spreading process on a range of networks containing different structures, we find that the speed and degree to which predictions of process outcomes – specifically initial growth rate and final size – are affected by out-degree truncation varies greatly. All processes are eventually predicted to have limited impact, however how much truncation is required varies. Notably, our ability to predict process outcomes is degraded more rapidly on stylized synthetic networks than on a set of empirical social contact networks from villages in Karnataka state, India.

Central to understanding the effect of out-degree truncation on predictions of spreading process outcomes is the transition when the network becomes fragmented and the size of the largest connected component rapidly decreases. In our analyses, the Power-Law and Degree-Assortative networks showed slow declines in predicted process outcomes as truncation increased, while the loss of accuracy was more rapid for both Triadic Clustering and Focal Clustering networks – which lost fidelity early on – and the Karnataka networks – which maintained fidelity for longer (Figure 3). The speed of initial growth was notably more variable for Degree-Assortative compared to all other network types for both no truncation and truncation at $2\langle k \rangle$, reflecting the importance of the initial infection sites when networks contain both highly and lowly connected regions. This variation in findings suggests that knowledge of the structure of a network for which one wishes to predict process spread is crucial in determining the level of resources that should place into measuring the full extent of the network itself: locally clustered networks may require more contacts, while those with fat-tailed degree distributions may require fewer. Of course, knowing the mean out-degree of a network is a pre-requisite to determining the level of truncation that can be tolerated.

In contrast to our conjectures, in no case did truncation increase either speed or size of process spread. The impact of truncation in reducing the number of observed ties appeared to overwhelm all other processes, not least by affecting the network characteristics of the truncation networks: truncation at $\langle k \rangle$ led to the Degree-Assortative networks being entirely non-assortative and the Triadic Clustering and Focal Clustering networks displaying very limited clustering; only modularity appeared to be maintained or even increased as the FCD threshold was lowered – potentially because of the breakup of the network into increasingly numbers of unconnected components. Further investigation might find levels of truncation at which epidemics severity is over-estimated, but in practical terms our findings point to a consistent underestimate of speed and attack rate using data truncated by strength.

In addition to network-level outcomes, it is instructive to consider variability in outcomes at the individual level. While it is clear that individuals with higher out-degree are more likely to become infected, it is also likely that those with more-connected neighbors will become infected more often, since these connected neighbors are more likely to be infected in the first place. This association can be seen in Figure 4 for the Karnataka networks (and Supplementary Figure 3 for synthetic networks). Low degree individuals are unlikely to be infected regardless of how well-connected their neighbors are, but for our exemplar infection neighbor degree has little impact for those with own degree greater than ten (Figure 4B). As truncation increases – and has a disproportionate impact on ties dropped to higher-degree neighbors – individuals with lower mean degree neighbors are predicted to be infected less often than those with the same degree, but lower mean neighbor degree (Figure 4C and D). This effect is particularly visible at the common FCD value of $\langle k \rangle$. These findings highlight that not only can truncation impact population-level predictions of infection risk, but they may also differentially affect individual-level predictions.

There are several ways in which this analysis could be extended. First, it might be informative to consider unweighted, rather than weighted, truncation. Weighted truncation is likely to minimize mis-estimation of local spreading processes, since close-knit groups are likely to be maintained at the expense of a realistic picture of cross-community connections. Unweighted truncation, in contrast, is likely to reduce the speed of epidemic spread generally, but maintain weak ties that span structural holes in the network(Burt, 2004). Second, one could investigate spreading processes based on edge weights, or using unit infectivity. Third, it might be worthwhile to run these analyses for a wide range of truncation levels, in order to evaluate which networks have more or less rapid transitions from relatively accurate epidemic predictions to relatively inaccurate ones, and at what level of truncation these transitions occur. Such an analysis would be particularly useful in the context of a specific empirical network and spreading process, rather than in the theoretical cases presented in this paper, as a precursor to the conduct of data collection in a survey. While we have used a range of network structures and a standard spreading process, our results are limited to the cases we have considered, and thus investigation of other structures and processes might be worthwhile.

The ultimate goal of our analysis is to arrive at more accurate predictions of process outcomes in the context of truncated contact data, the type of data that are common in the study of infectious diseases and public health interventions. In addition to our simulation approach, there is the potential for analytic work to evaluate the level of mis-prediction likely to arise under a given level of degree truncation, for given network structures. Ultimately, this should allow for us to adjust predictions for truncation. Such an approach might use statistical or mechanistic network models to simulate full networks congruent with both the estimated rate of truncation, and observed characteristics of the truncated network; simulations could then be run on these simulated networks to predict process outcomes. As noted above, although we have framed out-degree truncation here as resulting from the adoption of FCD, our methods are agnostic to the cause of truncation. Consequently our results may generalize to settings where some other mechanism, such as social stigma in the case of self-reported sexual networks, might lead to out-degree truncation.

Finally, there has been increasing research activity in the past few years into digitally mediated social networks, such as those resulting from mobile phone call and communication patterns. Reliance on these types of data requires the investigator to specify the width of the data aggregation window and other similar parameters, leading to effective degree truncation that is similar in its effects to the truncation resulting here from study design. It seems plausible that some of the insights we have obtained here, as well as some of our methods, could be translated to this research context.

**Conclusion**

We have shown, via simulation, that truncation of a network via FCD has a systematic impact on how processes are predicted to spread across this network. However, the degree of impact varies strongly by the level of truncation, and we find that the transition level – at which impact on predicted process outcomes shifts from small to considerable – varies by network structure. Supplementary information on the structure of the full network – potentially estimated from past egocentric or sociocentric studies in the same or similar populations – will thus often be crucial for increasing the accuracy of predictions of process spread for truncated network data.

**Supplementary material**

Supplementary Table 1: Descriptive statistics for the calibrated network graphs (mean and 95% confidence intervals)

Supplementary Table 2: Percentage of epidemic simulation runs infecting at least 10% of the population

Supplementary Figure 1: Time to infection of 10% of all individuals on networks, amongst epidemic simulation runs infecting at least 10% of the population

Supplementary Figure 2: Epidemic attack rate on networks, amongst epidemic simulation runs infecting at least 10% of the population.

Supplementary Figure 3: Mean neighbor degree vs. own degree for full and truncated synthetic networks

**Table 1: Target characteristic values for calibrated synthetic networks**

|  | Minimum | Median | Maximum |
|---|---|---|---|
| Degree-assortativity coefficient (r) | 0.283 | 0.421 | 0.797 |
| Triadic clustering coefficient (c) | 0.249 | 0.284 | 0.353 |
| Focal clustering coefficient (t) | 0.163 | 0.249 | 0.326 |
| Power-law degree exponent ($\gamma$) | -3.0 | -2.5 | -2.0 |

**Table 2: Population-level outcomes amongst epidemic simulation runs infecting at least 10% of the population**

|  | No truncation | | Truncation at twice mean degree | | Truncation at mean degree | | Truncation at half mean degree | |
|---|---|---|---|---|---|---|---|---|
| **Time to infection of 10% of population** | | | | | | | | |
| Degree-Assortative | 35.0 | [20.0 - 85.0] | 51.0 | [27.9 - 120.9] | 119.9 | [67.1 - 185.0] | 138.0 | [81.0 - 188.0] |
| Triadic Clustering | 17.0 | [12.0 - 27.0] | 22.0 | [15.0 - 34.0] | 61.0 | [36.0 - 127.0] | | |
| Focal Clustering | 18.0 | [11.0 - 39.0] | 32.0 | [20.0 - 65.0] | 96.9 | [51.9 - 174.4] | | |
| Power-Law | 8.0 | [5.0 - 19.0] | 16.9 | [9.0 - 38.0] | 40.0 | [17.9 - 107.1] | 72.9 | [35.0 - 153.1] |
| Karnataka villages | 15.0 | [9.0 - 27.0] | 21.0 | [12.3 - 40.0] | 43.0 | [23.0 - 100.9] | 88.4 | [39.0 - 175.4] |
| **Percentage of all individuals ever infectious** | | | | | | | | |
| Degree-Assortative | 46.6 | [39.3 - 52.8] | 39.5 | [27.4 - 47.2] | 15.2 | [10.4 - 26.4] | 11.5 | [10.2 - 16.6] |
| Triadic Clustering | 85.8 | [83.4 - 87.9] | 84.4 | [81.6 - 86.7] | 41.8 | [18.8 - 54.1] | | |
| Focal Clustering | 60.2 | [55.0 - 65.0] | 58.0 | [51.0 - 63.2] | 15.7 | [10.5 - 27.4] | | |
| Power-Law | 58.8 | [51.5 - 65.1] | 41.1 | [32.6 - 48.2] | 22.2 | [12.6 - 30.0] | 15.9 | [10.6 - 27.5] |
| Karnataka villages | 78.1 | [68.9 - 83.9] | 76.2 | [65.6 - 82.9] | 57.5 | [20.1 - 70.9] | 13.9 | [10.3 - 24.2] |
| **Percentage of 47,500 epidemics infecting at least 10% of the population** | 96.5 | | 93.1 | | 66.0 | | 7.1 | |

Figures show mean and 95% ranges for all runs from 10,000 simulations (7,500 for Karnataka villages) for which at least of 10% of individuals were ever infected. Note that the proportion of retained networks falls as the level of truncation rises (see Supplementary Table 2 for details); empty cells represent simulation types where no runs reached the 10% threshold. All network structures are those with highest network properties in each category (see Methods and Table 1).

**Figure 1: Schematic of study methodology**

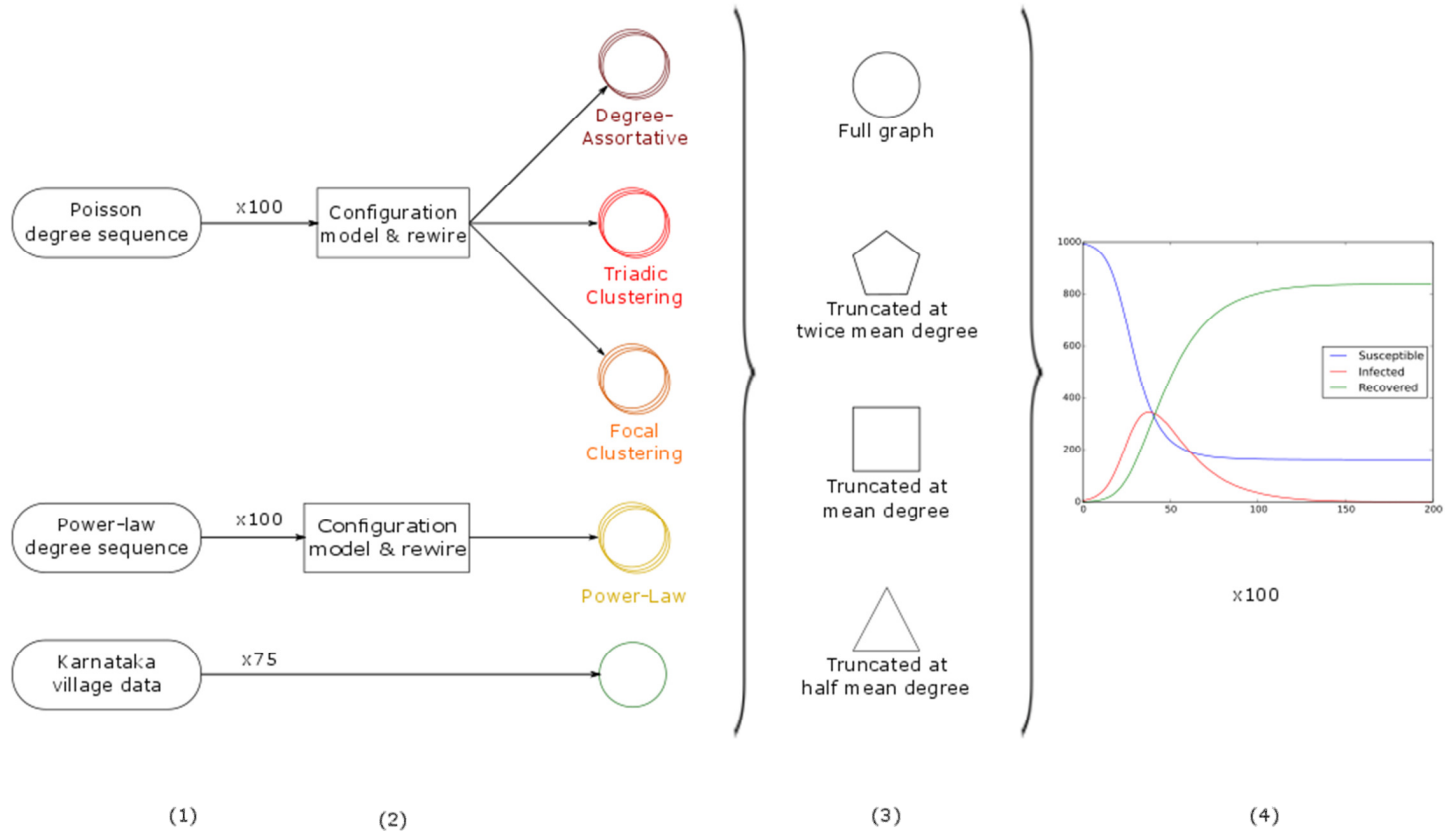

(1) For synthetic networks, 100 degree sequences were were generated. For the Karnataka village data, 75 empirical village datasets were used, and step 2 skipped. (2) Each degree sequence was converted into a network graph using the configuration model, and then each synthetic graph was calibrated based on target network values. (3) All networks were truncated at twice mean, mean and half mean degree. (4) 100 epidemics were run across each full and truncated network.

**Figure 2: Toy examples of truncation process for different synthetic graphs**

| Power-Law degree distribution | Degree-Assortative | Triadic Clustering |
|---|---|---|
| 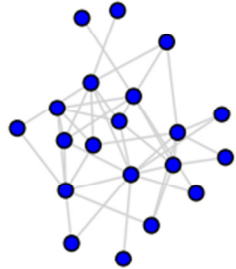 | 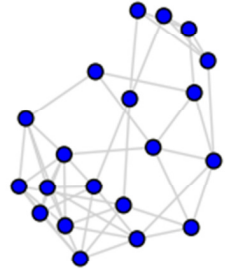 | 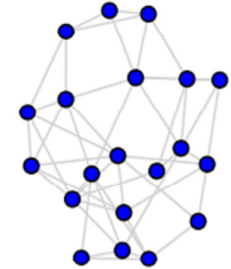 |
| 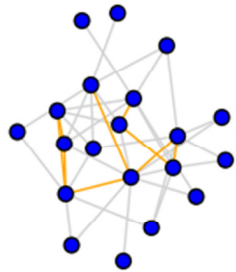 | 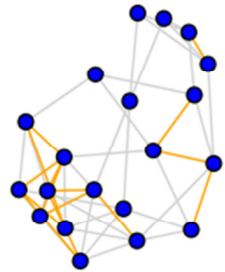 | 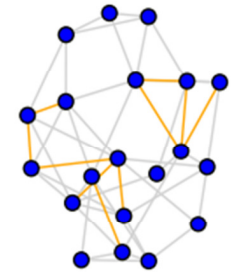 |

This figure shows three graphs each containing 20 nodes and with a mean degree of approximately 5. Each was generated by calibrating a configuration-generated graph through rewiring to achieve specific target values of different network characteristics. The top row shows each calibrated graph with edges marked in grey; the bottom row superimposes in orange the edges removed by truncating by tie strength at an out-degree of 3.

**Figure 3: Epidemic outcomes for simulation runs infecting at least 10% of the population across six network structures**

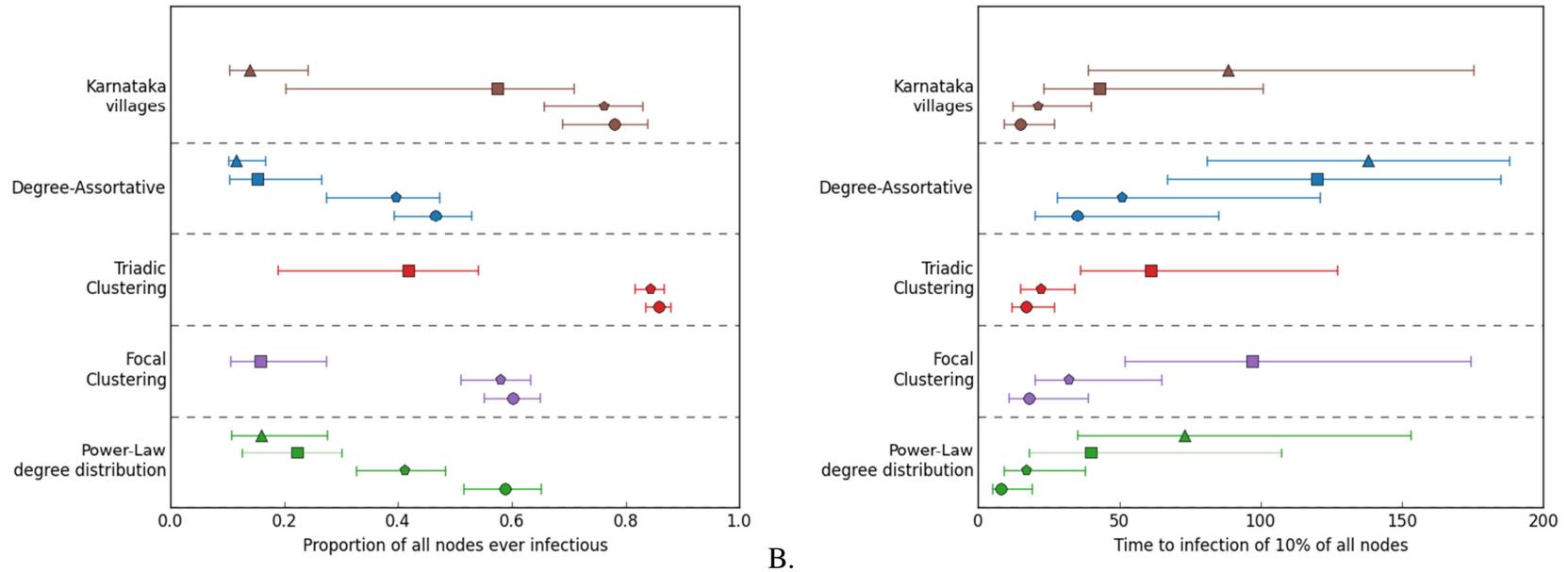

A.   B.

Figures show mean and 95% ranges for all runs from 10,000 simulations (7,500 for Karnataka villages) for which at least of 10% of individuals were ever infected. Simulation types are defined by out-degree truncation (Circles: no truncation; Hexagons: truncation at twice mean degree; Squares: truncation at mean degree; Triangles: truncation at half mean degree). All network structures are those with highest network properties in each category (see Methods and Table 1; full results for each network structure are available in Supplementary Figure 1 and Supplementary Figure 2). Empty lines represent simulation types where no runs reached the 10% threshold.

**Figure 4: Mean neighbor degree vs. own degree for full and truncated Karnataka village contact networks**

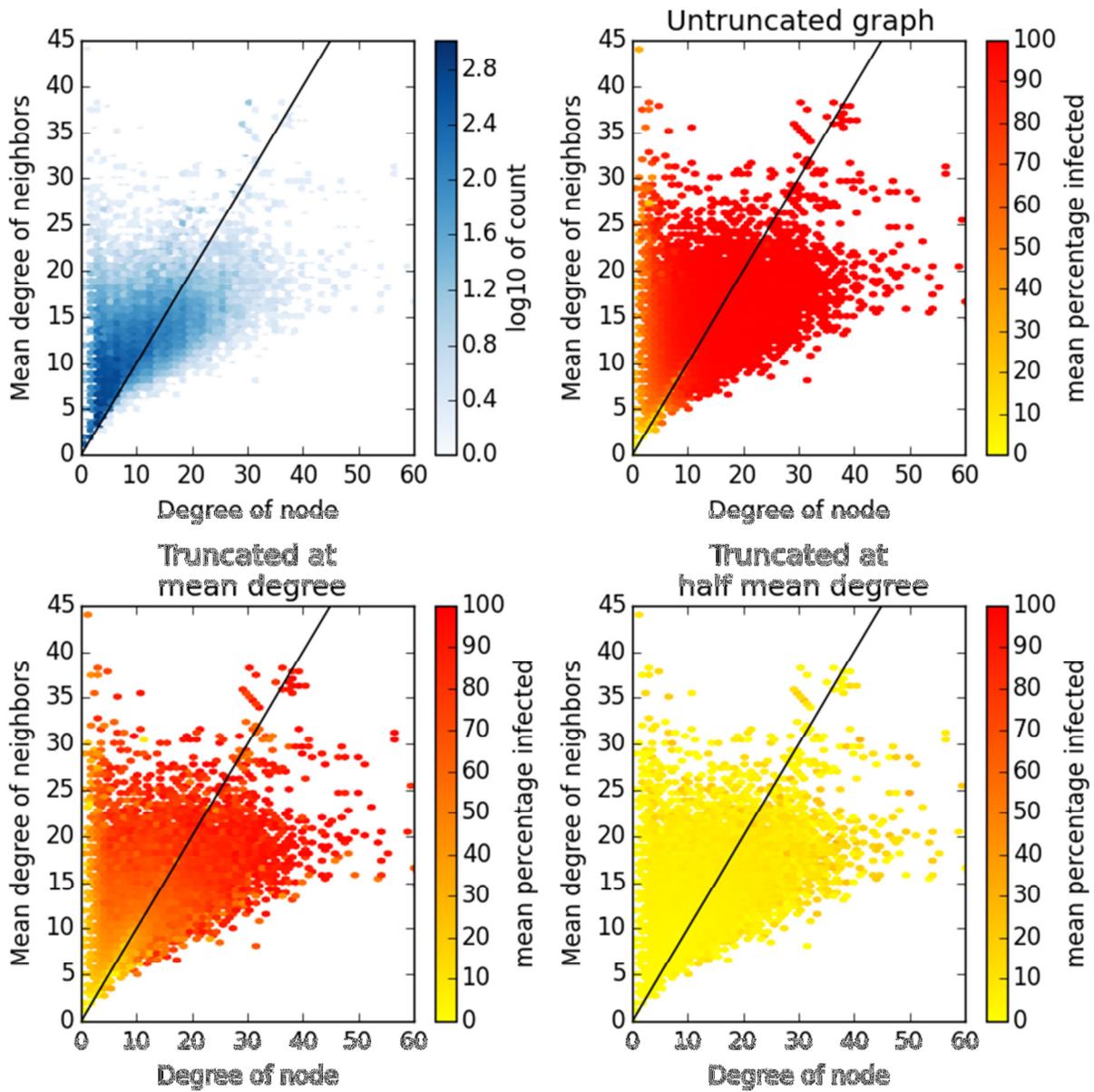

A. Density of ties in full graph (log-scale); B-D: Mean proportion of epidemic runs in which the node was infected (linear scale). The black diagonal line shows points of equal node and mean neighbor degree. In the full graph, most nodes are infected most of the time, except those with either very low degree or very low mean neighbor degree. When truncated at mean degree those with middling degree and mean neighbor degree are infected less often. When truncated at half mean degree almost no nodes are ever infected.

**Supplementary material**

**Title**: Impact of degree truncation on the spread of a contagious process on networks

**Supplementary Table 1: Descriptive statistics for the calibrated network graphs (mean and 95% confidence intervals)**

A. <u>Mean degree</u>

|  |  | Not truncated |  | Truncated at $2\langle k \rangle$ |  | Truncated at $\langle k \rangle$ |  | Truncated at $0.5\langle k \rangle$ |  |
|---|---|---|---|---|---|---|---|---|---|
| **Karnataka villages** |  | 8.39 | [7.84 - 8.97] | 7.21 | [6.72 - 7.60] | 5.54 | [4.77 - 5.65] | 3.90 | [2.78 - 3.95] |
| **Synthetic networks**: |  |  |  |  |  |  |  |  |  |
| Degree-Assortative | r = 0.283 | 7.86 | [7.86 - 7.86] | 7.68 | [7.67 - 7.68] | 5.74 | [5.72 - 5.76] | 3.22 | [3.20 - 3.24] |
|  | r = 0.421 | 7.86 | [7.86 - 7.86] | 7.64 | [7.63 - 7.65] | 5.67 | [5.65 - 5.69] | 3.16 | [3.13 - 3.18] |
|  | r = 0.797 | 7.86 | [7.86 - 7.86] | 7.54 | [7.53 - 7.55] | 5.40 | [5.38 - 5.42] | 2.93 | [2.91 - 2.95] |
| Triadic Clustering | c = 0.249 | 7.75 | [7.75 - 7.75] | 7.40 | [7.39 - 7.42] | 5.56 | [5.53 - 5.58] | 3.12 | [3.10 - 3.13] |
|  | c = 0.284 | 7.75 | [7.75 - 7.75] | 7.39 | [7.36 - 7.40] | 5.55 | [5.52 - 5.57] | 3.19 | [3.17 - 3.20] |
|  | c = 0.353 | 7.75 | [7.75 - 7.75] | 7.31 | [7.29 - 7.33] | 5.51 | [5.48 - 5.53] | 3.32 | [3.30 - 3.33] |
| Focal Clustering | t = 0.163 | 7.95 | [7.95 - 7.95] | 6.84 | [6.78 - 6.88] | 4.49 | [4.46 - 4.54] | 2.57 | [2.54 - 2.59] |
|  | t = 0.249 | 7.95 | [7.95 - 7.95] | 6.29 | [6.17 - 6.37] | 4.07 | [4.00 - 4.12] | 2.32 | [2.28 - 2.35] |
|  | t = 0.326 | 7.95 | [7.95 - 7.95] | 5.84 | [5.73 - 5.92] | 3.76 | [3.67 - 3.83] | 2.15 | [2.11 - 2.20] |
| Power-Law | $\gamma = 3$ | 7.78 | [7.66 - 7.83] | 6.58 | [6.50 - 6.63] | 4.70 | [4.66 - 4.74] | 2.89 | [2.87 - 2.91] |
|  | $\gamma = 2.5$ | 7.40 | [7.04 - 7.55] | 6.22 | [5.97 - 6.33] | 4.60 | [4.56 - 4.65] | 2.91 | [2.89 - 2.93] |
|  | $\gamma = 2$ | 6.18 | [5.89 - 6.46] | 4.78 | [4.44 - 5.02] | 4.00 | [3.51 - 4.18] | 2.88 | [2.85 - 2.91] |

B. <u>Degree-assortativity</u>

| | | Not truncated | | Truncated at $2\langle k \rangle$ | | Truncated at $\langle k \rangle$ | | Truncated at $0.5\langle k \rangle$ | |
|---|---|---|---|---|---|---|---|---|---|
| **Karnataka villages** | | 0.33 | [0.30 - 0.37] | 0.23 | [0.20 - 0.25] | 0.11 | [0.09 - 0.13] | 0.02 | [-0.02 - 0.05] |
| **Synthetic networks defined by**: | | | | | | | | | |
| Degree-Assortative | r = 0.283 | 0.28 | [0.28 - 0.28] | 0.25 | [0.25 - 0.26] | -0.02 | [-0.03 - -0.01] | -0.19 | [-0.20 - -0.18] |
| | r = 0.421 | 0.42 | [0.42 - 0.42] | 0.38 | [0.37 - 0.38] | -0.00 | [-0.01 - 0.01] | -0.19 | [-0.20 - -0.17] |
| | r = 0.797 | 0.80 | [0.80 - 0.80] | 0.69 | [0.68 - 0.69] | -0.00 | [-0.02 - 0.01] | -0.20 | [-0.21 - -0.18] |
| Triadic Clustering | c = 0.249 | -0.05 | [-0.06 - -0.04] | -0.10 | [-0.11 - -0.09] | -0.16 | [-0.17 - -0.15] | -0.25 | [-0.27 - -0.24] |
| | c = 0.284 | -0.05 | [-0.06 - -0.04] | -0.10 | [-0.11 - -0.09] | -0.17 | [-0.18 - -0.16] | -0.26 | [-0.27 - -0.25] |
| | c = 0.353 | -0.06 | [-0.07 - -0.05] | -0.11 | [-0.12 - -0.10] | -0.18 | [-0.19 - -0.17] | -0.27 | [-0.28 - -0.26] |
| Focal Clustering | t = 0.163 | 0.26 | [0.23 - 0.29] | 0.11 | [0.09 - 0.12] | -0.07 | [-0.08 - -0.06] | -0.18 | [-0.20 - -0.17] |
| | t = 0.249 | 0.50 | [0.46 - 0.55] | 0.12 | [0.11 - 0.14] | -0.10 | [-0.11 - -0.08] | -0.20 | [-0.22 - -0.19] |
| | t = 0.326 | 0.68 | [0.65 - 0.72] | 0.08 | [0.07 - 0.10] | -0.14 | [-0.15 - -0.13] | -0.23 | [-0.25 - -0.21] |
| Power-Law | γ = 3 | -0.04 | [-0.06 - -0.03] | -0.11 | [-0.13 - -0.09] | -0.12 | [-0.15 - -0.10] | -0.14 | [-0.18 - -0.10] |
| | γ = 2.5 | -0.10 | [-0.13 - -0.08] | -0.14 | [-0.16 - -0.12] | -0.14 | [-0.16 - -0.11] | -0.14 | [-0.16 - -0.12] |
| | γ = 2 | -0.22 | [-0.24 - -0.20] | -0.24 | [-0.26 - -0.21] | -0.23 | [-0.26 - -0.21] | -0.22 | [-0.25 - -0.20] |

C. <u>Modularity</u>

|  |  | Not truncated |  | Truncated at $2\langle k \rangle$ |  | Truncated at $\langle k \rangle$ |  | Truncated at $0.5\langle k \rangle$ |  |
|---|---|---|---|---|---|---|---|---|---|
| **Karnataka villages** |  | 0.79 | [0.77 - 0.82] | 0.81 | [0.79 - 0.84] | 0.84 | [0.82 - 0.86] | 0.87 | [0.84 - 0.90] |
| **Synthetic networks defined by**: |  |  |  |  |  |  |  |  |  |
| Degree-Assortative | r = 0.283 | 0.29 | [0.29 - 0.29] | 0.30 | [0.30 - 0.30] | 0.40 | [0.40 - 0.41] | 0.66 | [0.65 - 0.66] |
|  | r = 0.421 | 0.28 | [0.28 - 0.29] | 0.30 | [0.30 - 0.30] | 0.41 | [0.40 - 0.41] | 0.66 | [0.66 - 0.67] |
|  | r = 0.797 | 0.28 | [0.28 - 0.28] | 0.30 | [0.30 - 0.30] | 0.44 | [0.43 - 0.45] | 0.71 | [0.71 - 0.72] |
| Triadic Clustering | c = 0.249 | 0.46 | [0.45 - 0.46] | 0.46 | [0.45 - 0.46] | 0.48 | [0.48 - 0.49] | 0.68 | [0.67 - 0.68] |
|  | c = 0.284 | 0.47 | [0.47 - 0.48] | 0.47 | [0.47 - 0.48] | 0.49 | [0.49 - 0.50] | 0.67 | [0.67 - 0.68] |
|  | c = 0.353 | 0.50 | [0.49 - 0.50] | 0.50 | [0.49 - 0.50] | 0.52 | [0.51 - 0.52] | 0.66 | [0.66 - 0.67] |
| Focal Clustering | t = 0.163 | 0.66 | [0.65 - 0.67] | 0.62 | [0.61 - 0.63] | 0.60 | [0.59 - 0.60] | 0.76 | [0.76 - 0.77] |
|  | t = 0.249 | 0.82 | [0.81 - 0.83] | 0.78 | [0.77 - 0.79] | 0.72 | [0.72 - 0.74] | 0.81 | [0.81 - 0.82] |
|  | t = 0.326 | 0.90 | [0.89 - 0.91] | 0.87 | [0.86 - 0.89] | 0.83 | [0.81 - 0.84] | 0.86 | [0.85 - 0.87] |
| Power-Law | $\gamma = 3$ | 0.36 | [0.36 - 0.36] | 0.32 | [0.31 - 0.32] | 0.43 | [0.43 - 0.44] | 0.68 | [0.67 - 0.69] |
|  | $\gamma = 2.5$ | 0.36 | [0.35 - 0.36] | 0.34 | [0.33 - 0.35] | 0.45 | [0.45 - 0.46] | 0.68 | [0.67 - 0.68] |
|  | $\gamma = 2$ | 0.37 | [0.36 - 0.38] | 0.43 | [0.41 - 0.45] | 0.50 | [0.49 - 0.56] | 0.68 | [0.67 - 0.68] |

D. Triadic clustering coefficient

|  |  | Not truncated |  | Truncated at 2⟨k⟩ |  | Truncated at ⟨k⟩ |  | Truncated at 0.5⟨k⟩ |  |
| --- | --- | --- | --- | --- | --- | --- | --- | --- | --- |
| **Karnataka villages** |  | 0.64 | [0.63 - 0.66] | 0.60 | [0.57 - 0.61] | 0.50 | [0.48 - 0.51] | 0.34 | [0.27 - 0.37] |
| **Synthetic networks defined by**: |  |  |  |  |  |  |  |  |  |
| Degree-Assortative | r = 0.283 | 0.01 | [0.01 - 0.01] | 0.01 | [0.01 - 0.01] | 0.00 | [0.00 - 0.01] | 0.00 | [0.00 - 0.00] |
|  | r = 0.421 | 0.01 | [0.01 - 0.01] | 0.01 | [0.01 - 0.01] | 0.00 | [0.00 - 0.01] | 0.00 | [0.00 - 0.00] |
|  | r = 0.797 | 0.01 | [0.01 - 0.01] | 0.01 | [0.01 - 0.01] | 0.01 | [0.01 - 0.01] | 0.00 | [0.00 - 0.00] |
| Triadic Clustering | c = 0.249 | 0.29 | [0.29 - 0.30] | 0.26 | [0.26 - 0.26] | 0.13 | [0.12 - 0.13] | 0.03 | [0.03 - 0.04] |
|  | c = 0.284 | 0.34 | [0.34 - 0.34] | 0.30 | [0.29 - 0.30] | 0.15 | [0.15 - 0.16] | 0.04 | [0.04 - 0.05] |
|  | c = 0.353 | 0.43 | [0.43 - 0.43] | 0.37 | [0.36 - 0.37] | 0.20 | [0.19 - 0.20] | 0.07 | [0.06 - 0.07] |
| Focal Clustering | t = 0.163 | 0.37 | [0.37 - 0.38] | 0.28 | [0.27 - 0.28] | 0.12 | [0.12 - 0.13] | 0.04 | [0.04 - 0.05] |
|  | t = 0.249 | 0.43 | [0.42 - 0.44] | 0.30 | [0.29 - 0.31] | 0.15 | [0.13 - 0.15] | 0.06 | [0.05 - 0.06] |
|  | t = 0.326 | 0.45 | [0.44 - 0.46] | 0.31 | [0.30 - 0.32] | 0.16 | [0.15 - 0.17] | 0.06 | [0.05 - 0.07] |
| Power-Law | $\gamma = 3$ | 0.04 | [0.03 - 0.05] | 0.02 | [0.02 - 0.02] | 0.01 | [0.01 - 0.01] | 0.00 | [0.00 - 0.01] |
|  | $\gamma = 2.5$ | 0.09 | [0.07 - 0.13] | 0.04 | [0.03 - 0.05] | 0.02 | [0.02 - 0.03] | 0.01 | [0.01 - 0.01] |
|  | $\gamma = 2$ | 0.21 | [0.19 - 0.22] | 0.05 | [0.04 - 0.06] | 0.03 | [0.03 - 0.03] | 0.02 | [0.01 - 0.02] |

E. <u>Focal clustering coefficient</u>

|  |  | Not truncated |  | Truncated at 2⟨k⟩ |  | Truncated at ⟨k⟩ |  | Truncated at 0.5⟨k⟩ |  |
| --- | --- | --- | --- | --- | --- | --- | --- | --- | --- |
| **Karnataka villages** |  | 0.19 | [0.17 - 0.21] | 0.18 | [0.16 - 0.19] | 0.16 | [0.15 - 0.17] | 0.11 | [0.08 - 0.12] |
| **Synthetic networks defined by:** |  |  |  |  |  |  |  |  |  |
| Degree-Assortative | r = 0.283 | 0.00 | [0.00 - 0.00] | 0.00 | [0.00 - 0.00] | 0.00 | [0.00 - 0.00] | 0.00 | [0.00 - 0.00] |
|  | r = 0.421 | 0.01 | [0.00 - 0.01] | 0.00 | [0.00 - 0.00] | 0.00 | [0.00 - 0.00] | 0.00 | [0.00 - 0.00] |
|  | r = 0.797 | 0.01 | [0.01 - 0.01] | 0.01 | [0.01 - 0.01] | 0.00 | [0.00 - 0.00] | 0.00 | [0.00 - 0.00] |
| Triadic Clustering | c = 0.249 | 0.07 | [0.07 - 0.07] | 0.06 | [0.06 - 0.06] | 0.03 | [0.03 - 0.03] | 0.01 | [0.01 - 0.01] |
|  | c = 0.284 | 0.08 | [0.08 - 0.08] | 0.07 | [0.07 - 0.07] | 0.03 | [0.03 - 0.03] | 0.01 | [0.01 - 0.01] |
|  | c = 0.353 | 0.09 | [0.08 - 0.09] | 0.07 | [0.07 - 0.07] | 0.04 | [0.04 - 0.04] | 0.01 | [0.01 - 0.01] |
| Focal Clustering | t = 0.163 | 0.16 | [0.16 - 0.16] | 0.11 | [0.10 - 0.11] | 0.05 | [0.04 - 0.05] | 0.01 | [0.01 - 0.02] |
|  | t = 0.249 | 0.25 | [0.25 - 0.25] | 0.14 | [0.13 - 0.14] | 0.06 | [0.06 - 0.06] | 0.02 | [0.02 - 0.02] |
|  | t = 0.326 | 0.33 | [0.33 - 0.33] | 0.15 | [0.15 - 0.16] | 0.07 | [0.06 - 0.07] | 0.02 | [0.02 - 0.03] |
| Power-Law | $\gamma = 3$ | 0.02 | [0.02 - 0.02] | 0.01 | [0.01 - 0.01] | 0.00 | [0.00 - 0.00] | 0.00 | [0.00 - 0.00] |
|  | $\gamma = 2.5$ | 0.03 | [0.02 - 0.03] | 0.01 | [0.01 - 0.01] | 0.00 | [0.00 - 0.00] | 0.00 | [0.00 - 0.00] |
|  | $\gamma = 2$ | 0.04 | [0.04 - 0.05] | 0.01 | [0.01 - 0.01] | 0.00 | [0.00 - 0.00] | 0.00 | [0.00 - 0.00] |

F. <u>Average shortest path in Largest Connected Component</u>

|  |  | Not truncated |  | Truncated at $2\langle k \rangle$ |  | Truncated at $\langle k \rangle$ |  | Truncated at $0.5\langle k \rangle$ |  |
|---|---|---|---|---|---|---|---|---|---|
| **Karnataka villages** |  | 4.10 | [3.89 - 4.36] | 4.43 | [4.19 - 4.68] | 5.30 | [5.00 - 5.82] | 7.09 | [6.56 - 9.23] |
| **Synthetic networks defined by**: |  |  |  |  |  |  |  |  |  |
| Degree-Assortative | r = 0.283 | 3.61 | [3.61 - 3.62] | 3.65 | [3.65 - 3.65] | 4.17 | [4.16 - 4.18] | 6.17 | [6.13 - 6.23] |
|  | r = 0.421 | 3.65 | [3.65 - 3.65] | 3.69 | [3.69 - 3.69] | 4.22 | [4.21 - 4.23] | 6.36 | [6.29 - 6.41] |
|  | r = 0.797 | 3.88 | [3.87 - 3.88] | 3.91 | [3.90 - 3.91] | 4.47 | [4.47 - 4.48] | 7.36 | [7.28 - 7.46] |
| Triadic Clustering | c = 0.249 | 3.71 | [3.70 - 3.72] | 3.78 | [3.77 - 3.79] | 4.22 | [4.20 - 4.23] | 6.35 | [6.28 - 6.42] |
|  | c = 0.284 | 3.70 | [3.70 - 3.72] | 3.78 | [3.77 - 3.79] | 4.21 | [4.20 - 4.23] | 6.11 | [6.05 - 6.17] |
|  | c = 0.353 | 3.69 | [3.68 - 3.70] | 3.78 | [3.77 - 3.79] | 4.20 | [4.18 - 4.22] | 5.75 | [5.70 - 5.80] |
| Focal Clustering | t = 0.163 | 4.09 | [4.07 - 4.12] | 4.21 | [4.18 - 4.23] | 4.91 | [4.88 - 4.94] | 7.94 | [7.84 - 8.07] |
|  | t = 0.249 | 4.61 | [4.56 - 4.66] | 4.73 | [4.68 - 4.78] | 5.39 | [5.34 - 5.45] | 8.33 | [8.26 - 8.47] |
|  | t = 0.326 | 5.23 | [5.10 - 5.39] | 5.34 | [5.20 - 5.51] | 5.98 | [5.83 - 6.17] | 8.85 | [8.60 - 9.14] |
| Power-Law | $\gamma = 3$ | 3.35 | [3.30 - 3.38] | 3.61 | [3.56 - 3.64] | 4.25 | [4.18 - 4.30] | 6.34 | [6.12 - 6.51] |
|  | $\gamma = 2.5$ | 3.16 | [3.09 - 3.23] | 3.43 | [3.36 - 3.51] | 3.93 | [3.79 - 4.06] | 5.52 | [5.22 - 5.80] |
|  | $\gamma = 2$ | 3.07 | [3.03 - 3.10] | 3.50 | [3.45 - 3.54] | 3.85 | [3.79 - 3.93] | 4.70 | [4.59 - 4.83] |

$\langle k \rangle$: Mean degree of nodes in a given graph. For definitions of $r$, c, $t$, $\gamma$ and $\delta$ and how they define each synthetic network type, please see main text of paper.

**Supplementary Table 2: Percentage of epidemic simulation runs infecting at least 10% of the population**

|  |  | Not truncated | Truncated at $2\langle k \rangle$ | Truncated at $\langle k \rangle$ | Truncated at $0.5\langle k \rangle$ |
|---|---|---|---|---|---|
| **Karnataka villages** |  | 99.5 | 99.3 | 90.4 | 11.9 |
| **Synthetic networks defined by**: |  |  |  |  |  |
| Degree-Assortative | r = 0.283 | 91.1 | 90.1 | 76.0 | 11.7 |
|  | r = 0.421 | 89.9 | 88.9 | 69.3 | 8.7 |
|  | r = 0.797 | 89.1 | 82.6 | 26.7 | 1.0 |
| Triadic Clustering | c = 0.249 | 99.8 | 99.8 | 87.3 | 0.0 |
|  | c = 0.284 | 99.9 | 99.8 | 92.1 | 0.0 |
|  | c = 0.353 | 99.8 | 99.8 | 95.9 | 0.0 |
| Focal Clustering | t = 0.163 | 99.6 | 99.4 | 55.6 | 0.0 |
|  | t = 0.249 | 98.9 | 98.3 | 66.0 | 0.0 |
|  | t = 0.326 | 97.5 | 96.4 | 66.7 | 0.0 |
| Power-Law | $\gamma = 3$ | 98.6 | 92.1 | 43.6 | 9.7 |
|  | $\gamma = 2.5$ | 98.9 | 95.1 | 51.9 | 15.0 |
|  | $\gamma = 2$ | 97.5 | 89.1 | 56.3 | 23.8 |

Figures are percentage points of 10,000 runs (synthetic networks) or 7500 runs (Karnataka villages).

**Supplementary Figure 1: Time to infection of 10% of all individuals on networks, amongst epidemic simulation runs infecting at least 10% of the population**

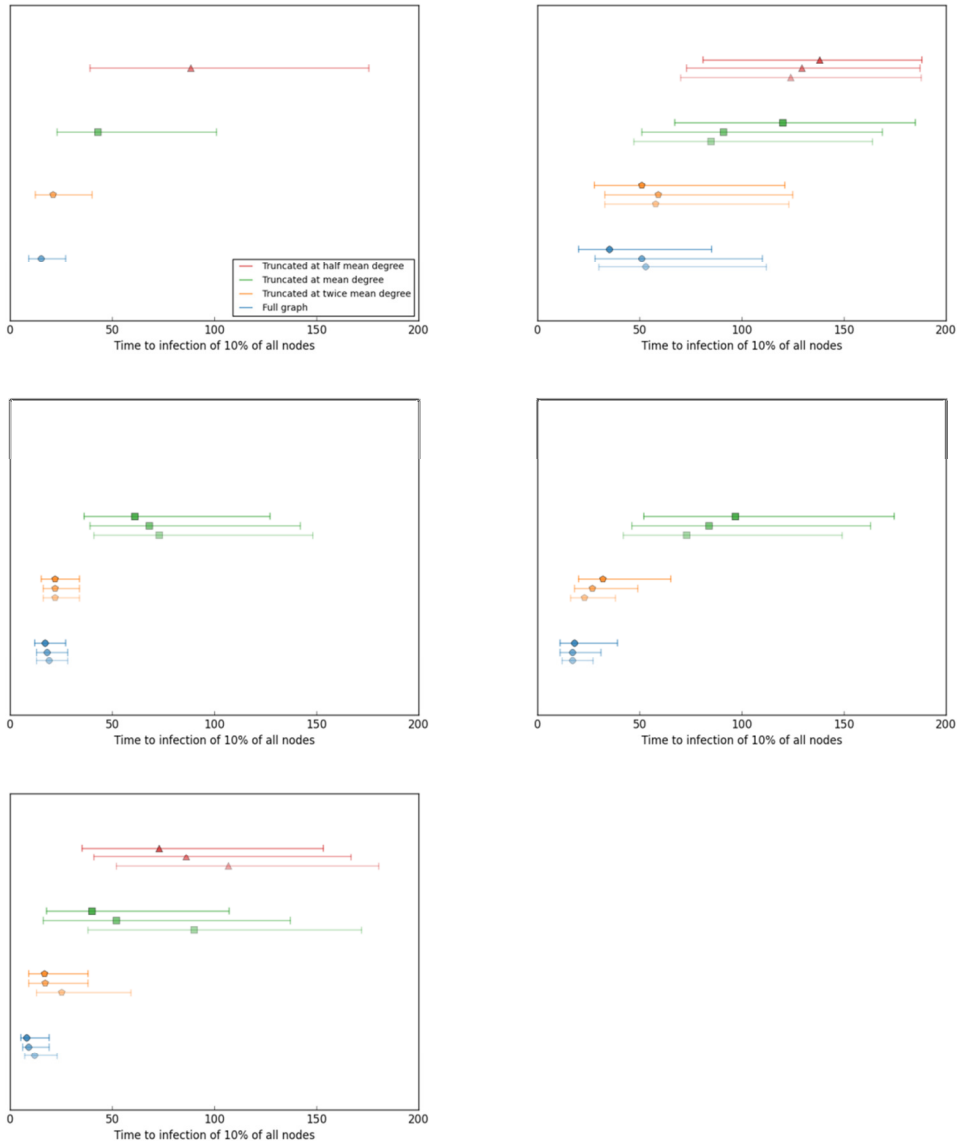

A: Karnataka villages; B: Degree-Assortative; C: Triadic Clustering; D: Focal Clustering; E: Power-Law networks. Figures show mean and 95% ranges for all runs from 10,000 simulations (7,500 for Karnataka villages) for which at least of 10% of individuals were ever infected. Simulation types are defined by truncation (see legend) and level of calibration – darker shading represents stronger calibration towards higher values of network properties (see Table 1). Empty lines represent simulation types where no runs reached the 10% threshold.

**Supplementary Figure 2: Epidemic attack rate on networks, amongst epidemic simulation runs infecting at least 10% of the population.**

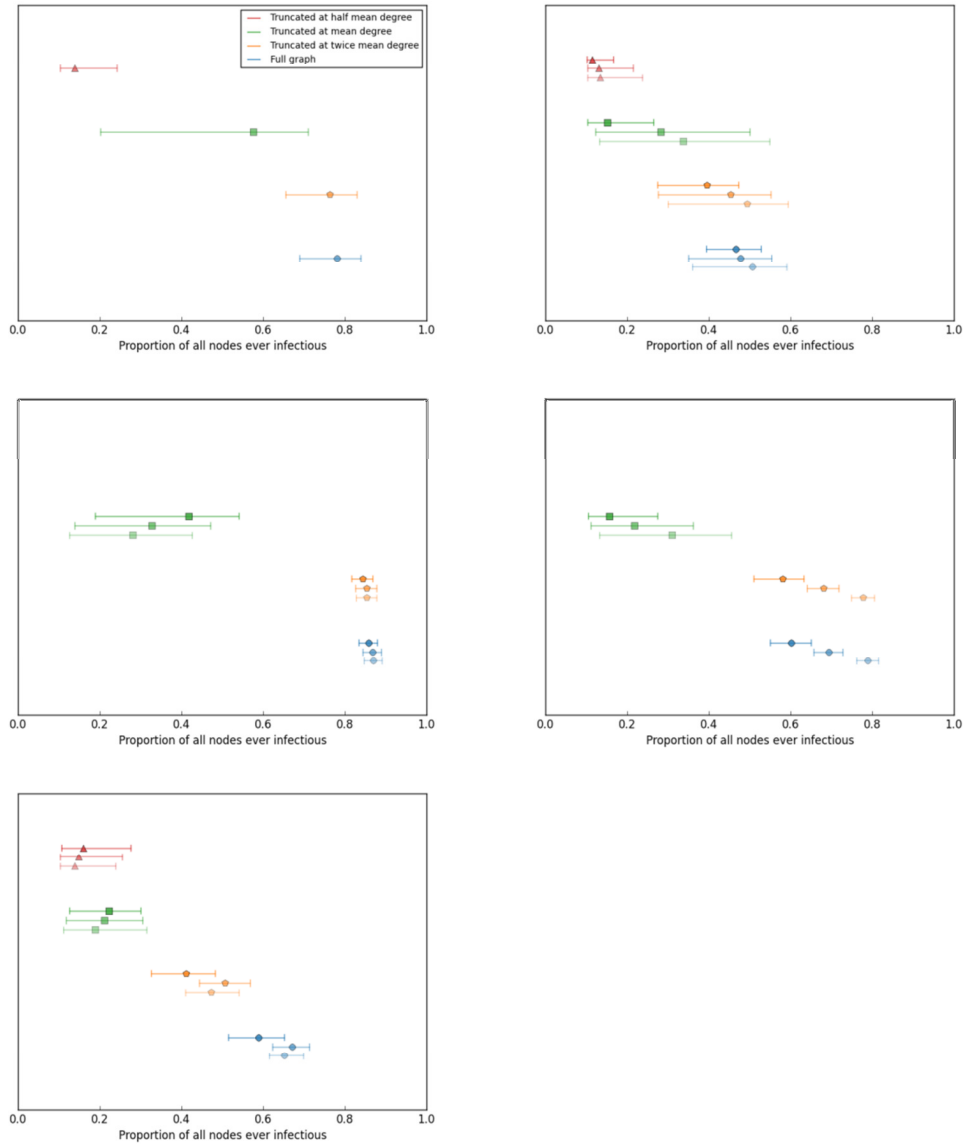

A: Karnataka villages; B: Degree-Assortative; C: Triadic Clustering; D: Focal Clustering; E: Power-Law networks. Figures show mean and 95% ranges for all runs from 10,000 simulations (7,500 for Karnataka villages) for which at least of 10% of individuals were ever infected. Simulation types are defined by truncation (see legend) and level of calibration – darker shading represents stronger calibration towards higher values of network properties (see Table 1). Empty lines represent simulation types where no runs reached the 10% threshold.

**Supplementary Figure 3: Mean neighbor degree vs. own degree for full and truncated synthetic networks**

For each set of figures below:

A. Full graph; B: graph truncated at twice mean degree; C: graph truncated at mean degree; D: graph truncated at half mean degree. Within each cell, darker=more: Blue (A1): Initial density of ties (log-scale); Green (B1, C1, D1): Mean proportion of neighbors dropped (linear scale); Red-Yellow (A2, B2, C2, D2): Mean proportion of epidemic runs in which the node was infected (linear scale). The black diagonal line shows points of equal node and mean neighbor degree.

I. <u>Degree-Assortative</u>

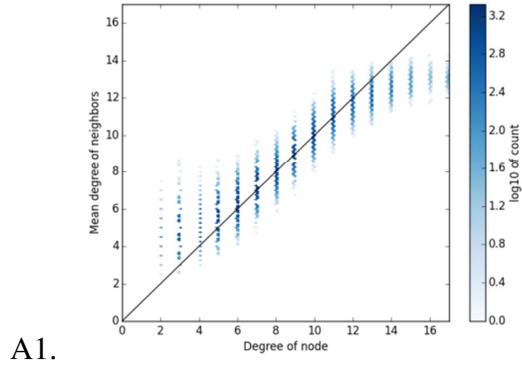

A1.

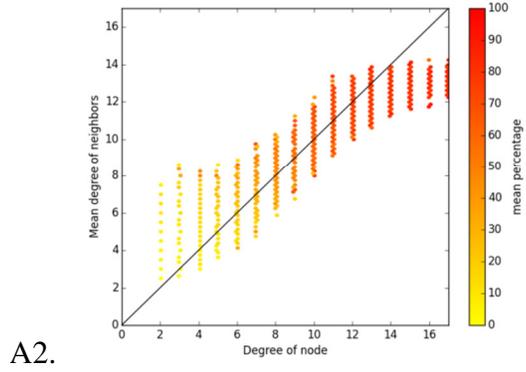

A2.

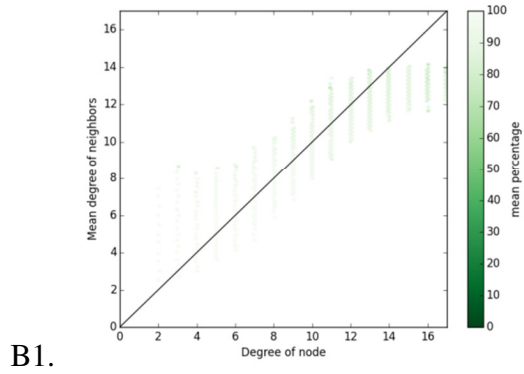

B1.

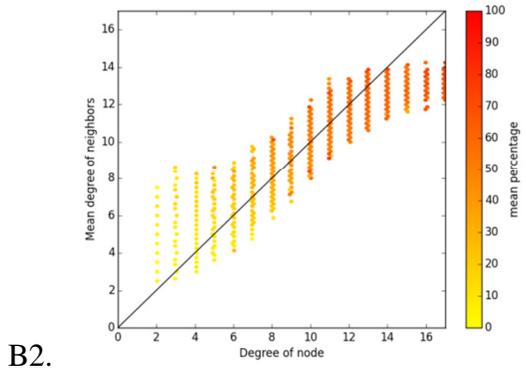

B2.

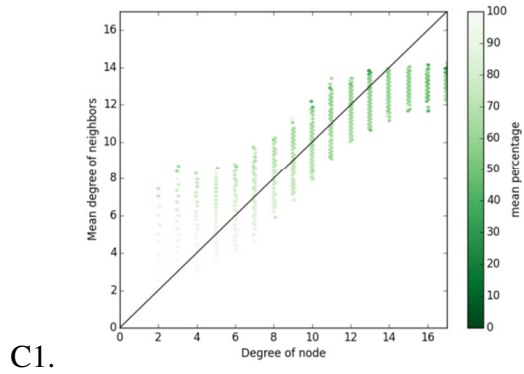

C1.

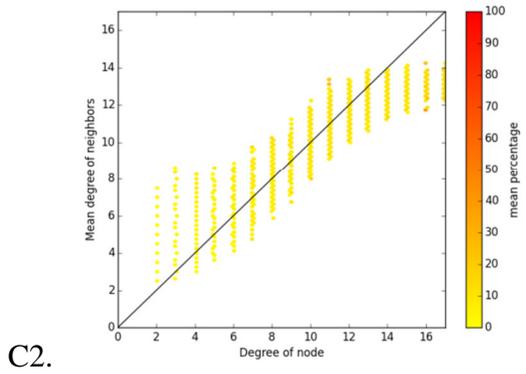

C2.

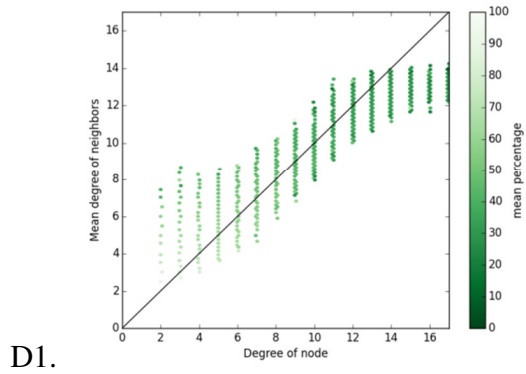

D1.

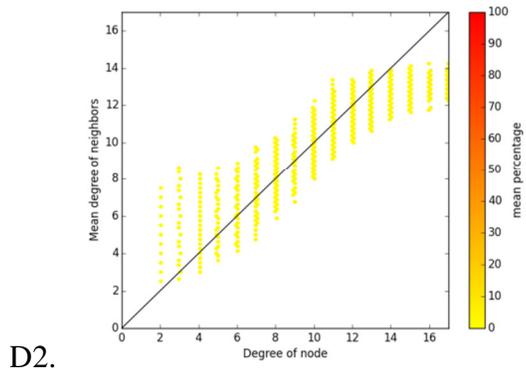

D2.

## II. Triadic Clustering

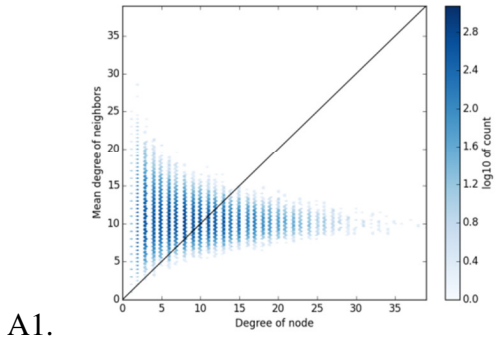 A1.

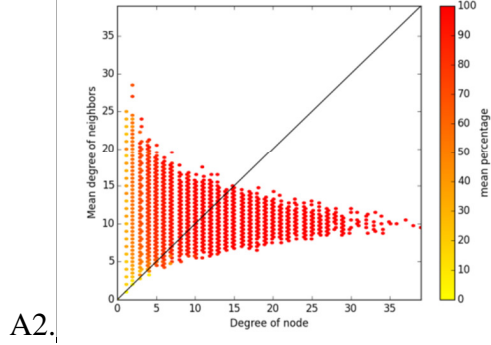 A2.

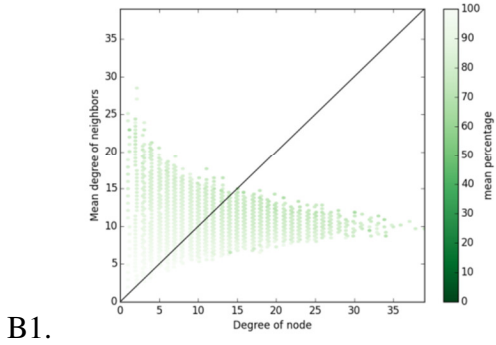 B1.

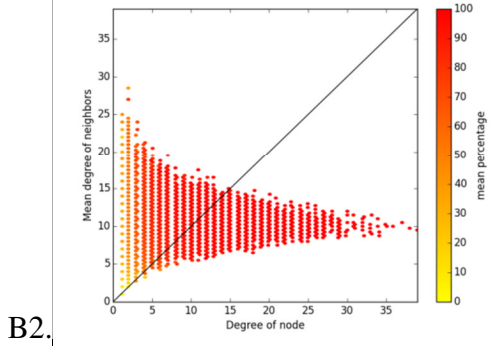 B2.

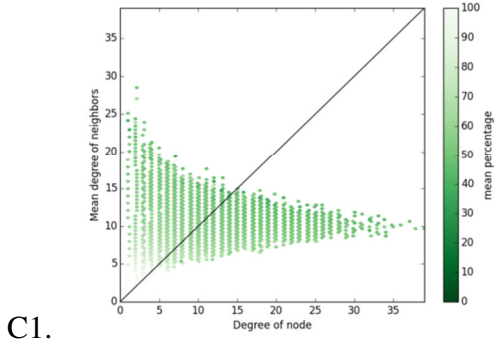 C1.

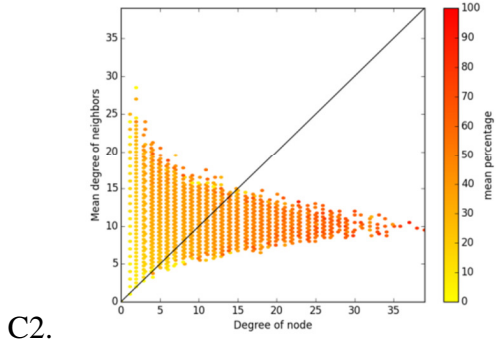 C2.

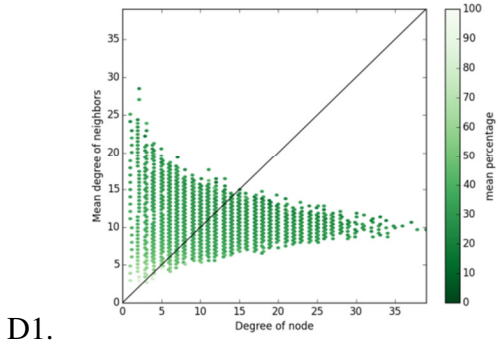 D1.

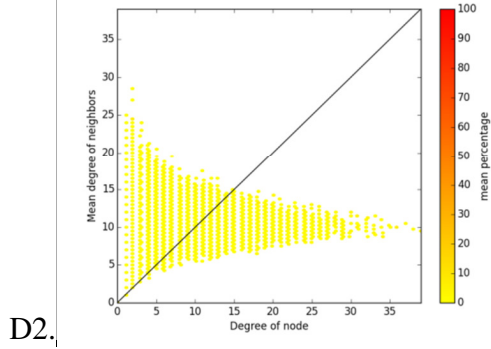 D2.

III.  <u>Focal Clustering</u>

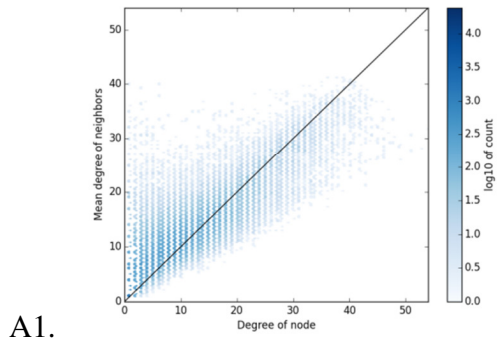

A1.

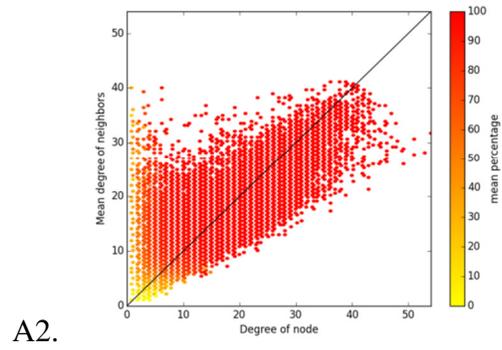

A2.

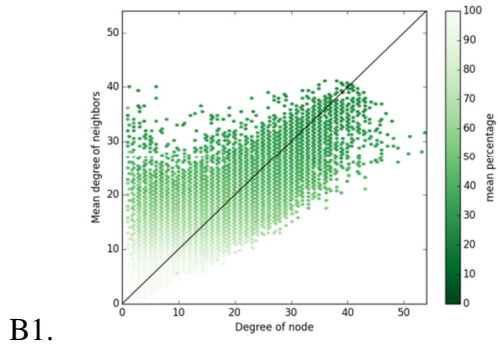

B1.

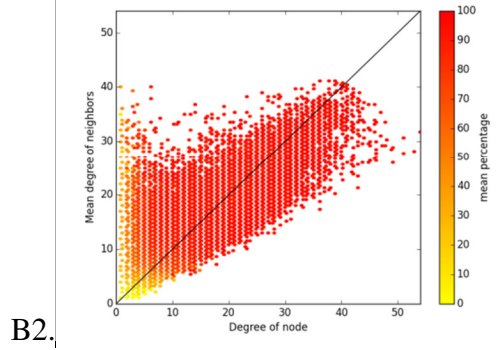

B2.

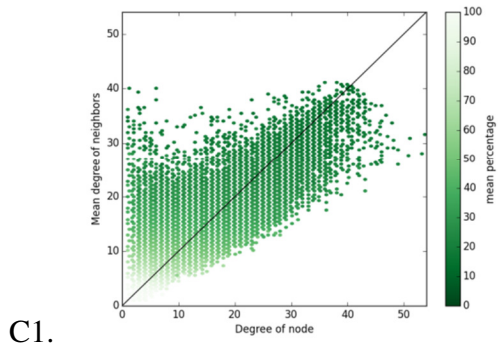

C1.

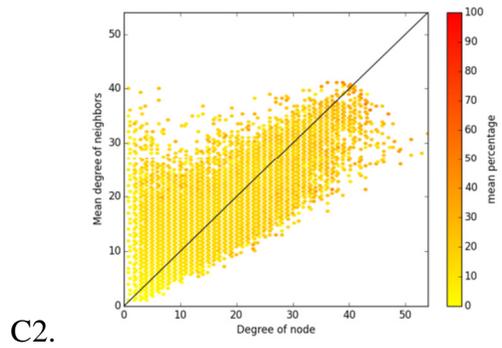

C2.

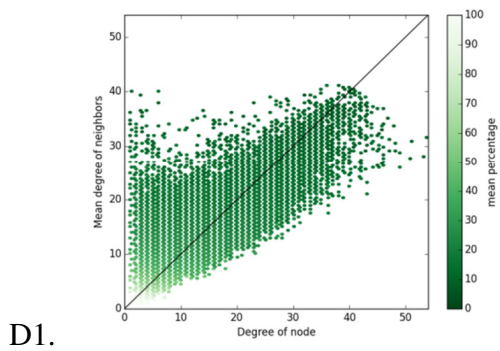

D1.

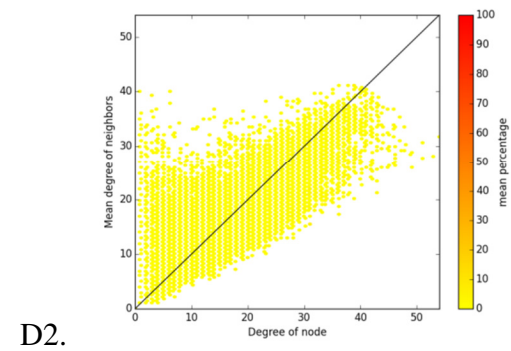

D2.

## IV. Power-Law degree distribution

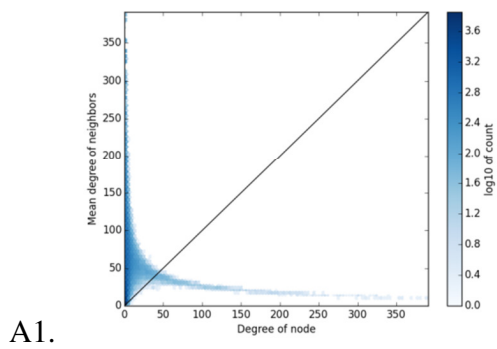
A1.

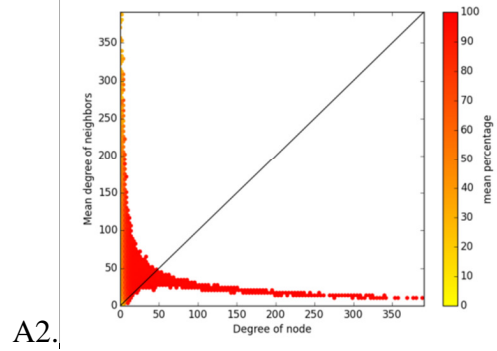
A2.

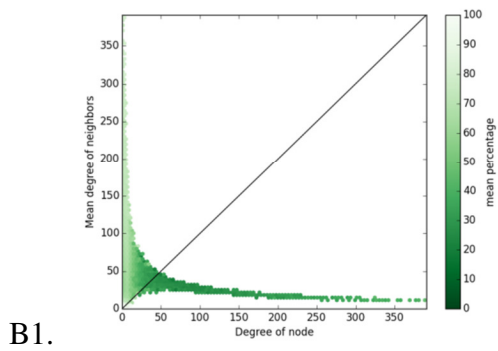
B1.

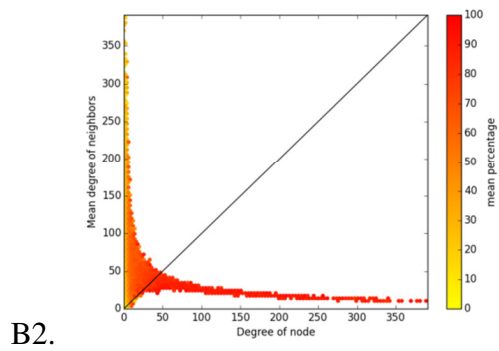
B2.

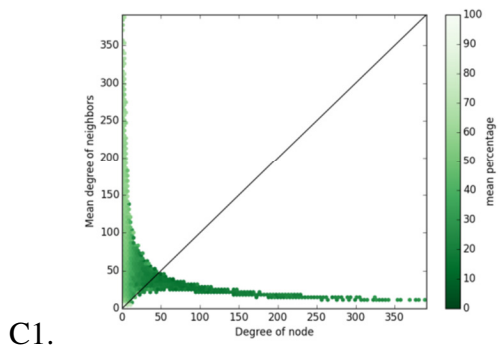
C1.

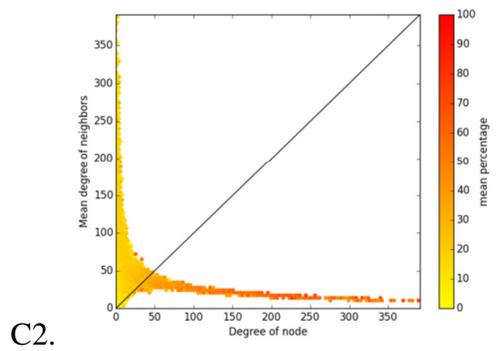
C2.

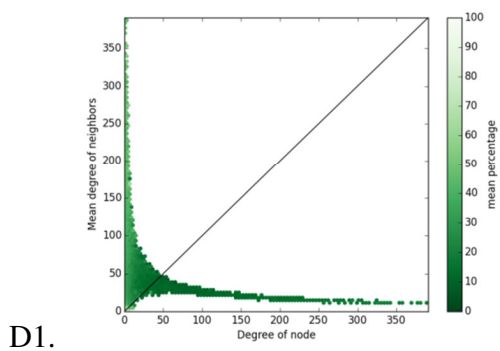
D1.

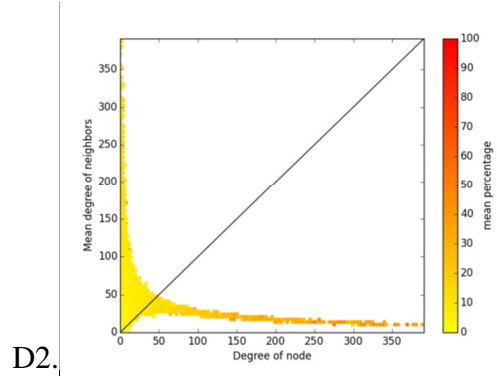
D2.

# V. Karnataka villages

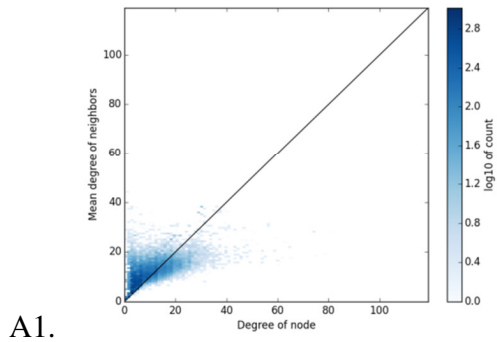

A1.

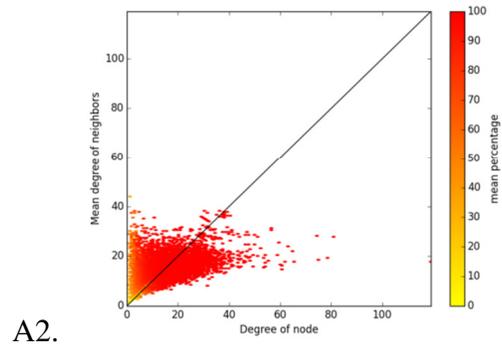

A2.

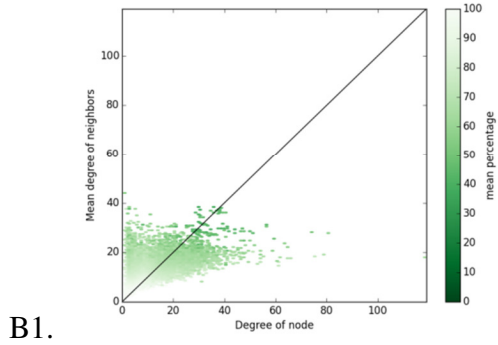

B1.

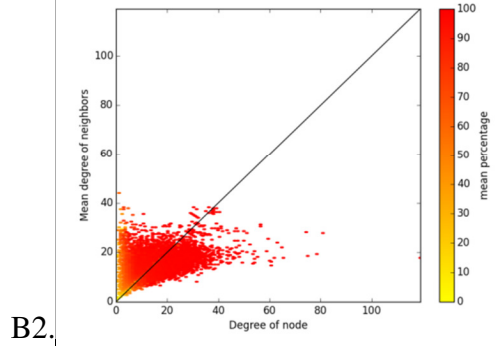

B2.

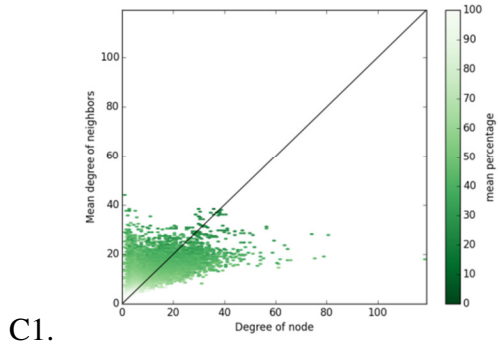

C1.

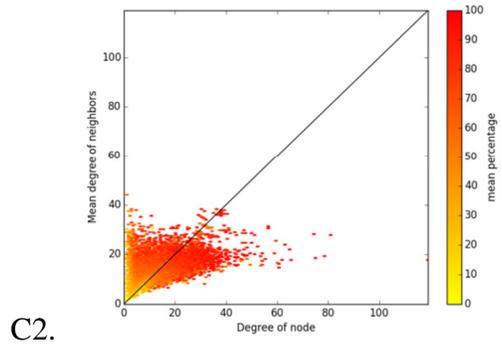

C2.

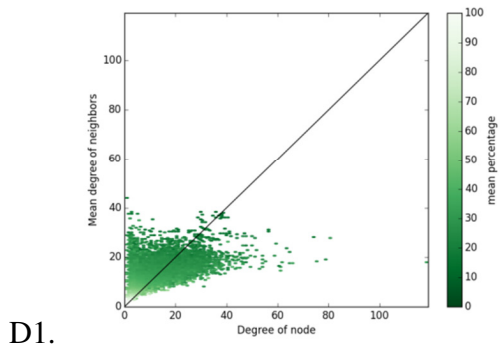

D1.

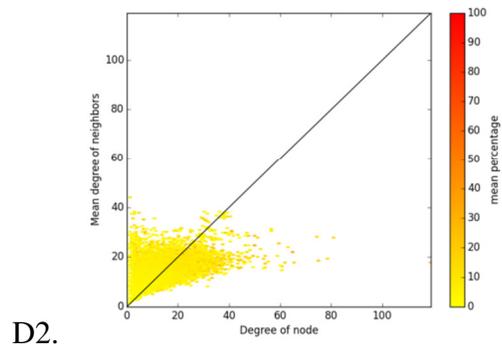

D2.